\documentclass[aps,prl,twocolumn,groupedaddress,superscriptaddress,preprintnumbers,amsmath,amssymb,nofootinbib,superscriptaddress,notitlepage]
{revtex4}

\usepackage{epsfig}
\usepackage[utf8]{inputenc}
\usepackage[dvipsnames]{xcolor}
\usepackage[normalem]{ulem}

\definecolor{blue}{HTML}{4169E1}
\definecolor{red}{HTML}{DC143C}
\definecolor{green}{HTML}{2E8B57}
\definecolor{violet}{HTML}{FF00D4}
\definecolor{mandarin}{HTML}{FF9933}

\newcommand{\eg}{\textit{e.g.}}
\newcommand{\ie}{\textit{i.e.}}

\newcommand{\ve}[1]{\ensuremath{\boldsymbol{#1}}}

\begin{document}

\title{Triple-X and beyond: hadronic systems of three and more X(3872)
}
\author{Lorenzo Contessi}\email{lorenzo@contessi.net}
\affiliation{
  IRFU, CEA, Université Paris-Saclay, 91191 Gif-sur-Yvette, France
}

\author{Johannes Kirscher}
\affiliation{
  Theoretical Physics Division, School of Physics and Astronomy,\\
  The University of Manchester, Manchester, M13 9PL, UK}
  
  \author{Manuel Pavon Valderrama}\email{mpavon@buaa.edu.cn}
\affiliation{School of Physics,
International Research Center for Nuclei and Particles in the Cosmos and \\
Beijing Key Laboratory of Advanced Nuclear Materials and Physics, \\
Beihang University, Beijing 100191, China}

\date{\today}

\begin{abstract} 
  \rule{0ex}{3ex}
  The $X(3872)$ resonance has been conjectured to be a $J^{PC} = 1^{++}$
  charm meson-antimeson two-body molecule.
  Meanwhile, there is no experimental evidence for larger, few-body
  compounds of multiple charm meson-antimeson pairs which would resemble larger
  molecules or nuclei.
  Here, we investigate such multi-meson states to
  the extent of what can be deduced theoretically from essentials of the interaction
  between uncharged $D^{0}$ and $D^{*0}$ mesons.
  From a molecular $X(3872)$,
  we predict a $4X$ ($4^{++}$) octamer with 
  a binding energy \mbox{$B_{4X} > 2.08\,{\rm MeV}$,}
  assuming a $D^{*0} \bar{D}^0$ system close to
  the unitary limit (as suggested by the mass of the $X(3872)$).
  If we consider heavy-quark spin symmetry explicitly,
  the $D^{*0} \bar{D}^{*0}$ ($2^{++}$) system is close to unitarity, too.
  In this case, we predict a bound $3X$ ($3^{++}$) hexamer
  with $B_{3X} > 2.29\,{\rm MeV}$
  and a more deeply bound $4X$ octamer 
  with $B_{4X} > 11.21\,{\rm MeV}$.
  These results exemplify with hadronic molecules a more general phenomenon of
  equal-mass two-species Bose systems comprised of equal number of either type:
  the emergence of unbound four- and six-boson clusters in
  the limit of a short-range two-body interaction which
  acts only between bosons of different species.
  Finally, we also study the conditions under which a $2X$ ($2^{++}$) tetramer
  might form.
\end{abstract}

\maketitle

Systems of particles with a two-body scattering length $a$
significantly larger than the interaction range $R$ \mbox{($a \gg R$)}
share a series of common/universal properties, which encompass a multitude of phenomena in atomic, nuclear,
and particle physics~\cite{Braaten:2004rn}.
This invariance with respect to a continuous scale transformation, however, holds strictly only
in the two-body sector.
In the few-body spectrum, this {\it continuous}
scale invariance 
survives only partially 
in a {\it discrete} version.
An example of this is the Efimov effect~\cite{Efimov:1970zz},
\ie~the appearance of a geometric bound-state spectrum
of three-boson systems in the unitary limit ($a/R \to \infty$).
This effect was found for the first time a decade ago
in experiments with caesium atoms~\cite{Kraemer:2006},
and it is now known to extend to systems of non-identical particles~\cite{Helfrich:2010yr}
as well as systems of more than three particles~\cite{Castin:2010,Bazak:2017}.
In nuclear physics, the Efimov effect plays a role
in the description of the triton~\cite{Bedaque:1998kg,Bedaque:1999ve} and
$^4{\rm He}$~\cite{Konig:2016utl}, halo nuclei~\cite{Federov:1994cf,Horiuchi:2006ds,Canham:2008jd,Acharya:2013aea,Ji:2014wta},
and the Hoyle state~\cite{Hammer:2008ra,Higa:2008dn}.
In bosonic systems with more than three particles, the same effect realizes
stable clusters (see \eg,~\cite{manybosons} where up to 60 bosons where analyzed).

Compared with atoms and nucleons,
it is more difficult to find instances of universality in hadronic physics
where
the $X(3872)$ resonance~\cite{Choi:2003ue}
might qualify as a hadronic system close to the unitary limit.
The $X$ has been conjectured to be a hadronic molecule~\cite{Tornqvist:2003na,Voloshin:2003nt}, more precisely, a 
relatively shallow bound state of two hadrons
because of its proximity to the $D^{*0} \bar{D}^0$ threshold
($\sim0.01~{\rm MeV}$) and its narrow width.
This shallowness, in particular, is a signature of universal
behaviour~\cite{Braaten:2003he}.
To explore the consequences of universality, we will describe $X$ and multi-$X$ systems with
a contact-range theory~\cite{Braaten:2003he} and use $D$ ($\bar{D}$) mesons (antimesons) as fundamental degrees of freedom.
With this approach, we aim to expose characteristic features of composite systems with the minimal assumptions and data on the constituents.
Alternative descriptions may improve on accuracy if the coupling to other channels, meson
exchanges, etc.~\cite{Swanson:2003tb,Gamermann:2009fv,Gamermann:2009uq,Lee:2009hy,Baru:2011rs}
is considered. The bulk properties of
the systems we analyse below, however, will not be affected by these refinements.

The identification of universal properties in systems composed of
more than two charm mesons is an intriguing open question because
charm meson-antimeson interactions produce qualitatively new features
that are absent in systems of identical particles.
For instance, both three-body systems $D^0 D^0 \bar{D}^{*0}$ and
$D^{*0} D^{*0} \bar{D}^{0}$ do neither form trimers nor do they display
the Efimov effect~\cite{Canham:2009zq}.
Along with heavy-quark spin symmetry (HQSS)~\cite{Isgur:1989ed,Isgur:1989vq}~and the associated
more tightly constrained
charm meson-antimeson potential enter new features.
In the two-body sector, we expect from HQSS
the interaction in the $D^{*0} \bar{D}^0$ ($J^{PC} = 1^{++}$) $X$-channel
to be identical to the one in the $D^{*0} \bar{D}^{*0}$ ($2^{++}$)
channel, suggesting the existence of a partner
molecule of the $X$~\cite{Valderrama:2012jv,Nieves:2012tt}. 
Like the $X(3872)$, this partner is expected to be shallow but its survival
as a bound or virtual state, or as a resonance depends on a number of
uncertainties~\cite{Nieves:2012tt,Baru:2016iwj,Cincioglu:2016fkm}.
HQSS challenges
the original expectation of an unbound $D^{*0} D^{*0} \bar{D}^0$ $J=2$ three-body system, and could thus facilitate the Efimov effect~\cite{Valderrama:2018sap}.

In four-meson systems and beyond, we expect to find
new universal phenomena different from the ones known to emerge in 
atomic and nuclear composites~\cite{Tjon,manybosons}.
We will consider, in particular, systems of $N=2,3,4$ $D^0 \bar{D}^{*0}$ pairs
with maximum spin, \ie~$J=2,3,4$, respectively.
Bound ``polymers'' of this kind
exhibit a characteristic
scaling inversely proportional to the square of
the interaction range, \ie~$B_{2 N} \propto 1/R^2$.
We infer from this scaling the Thomas collapse~\cite{Thomas:1935zz} of
these systems along with the implied Efimov effect.
This collapse is eventually avoided owing to short-range effects, \eg~the finite interaction range. Conversely, the Efimov effect is impaired by long-range deviations from unitarity, \ie~a finite scattering length.
Specifically, $D^{*0} \bar{D}^{*0}$ pairs can decay strongly to $D^{(*)0} \bar{D}^{0}$/$D^{(*)+} \bar{D}^{-}$ via a short-range $D$-wave
operator~\cite{Albaladejo:2015dsa} inducing such a finite interaction width.
Using data on the $X$ in support of the assumption of an
infinite $D^{*0} \bar{D}^{0}$ scattering length (zero binding energy of
the $X$ molecule) and disregarding HQSS,
we predict a bound state of four $X$'s: an octamer.
As the tetramer and the hexamer are unbound under these circumstances, this
octamer resembles a {\it so-called} Brunnian~\cite{Baas:2012za,brunni}~state:
a generalization of a Borromean structure.
Finally, we predict that a $D^{*0} \bar{D}^{*0}$
interaction close to the unitary limit will stabilize the hexamer and thus
induce the transition from a Brunnian to a Borromean system
(a still unbound tetramer with a hexamer resembling a Borromean
bound state of $X$'s).

\vspace{2mm}
\textbf{Theory and calculation method: }

\begin{figure}
\centering
\includegraphics[width=\linewidth]{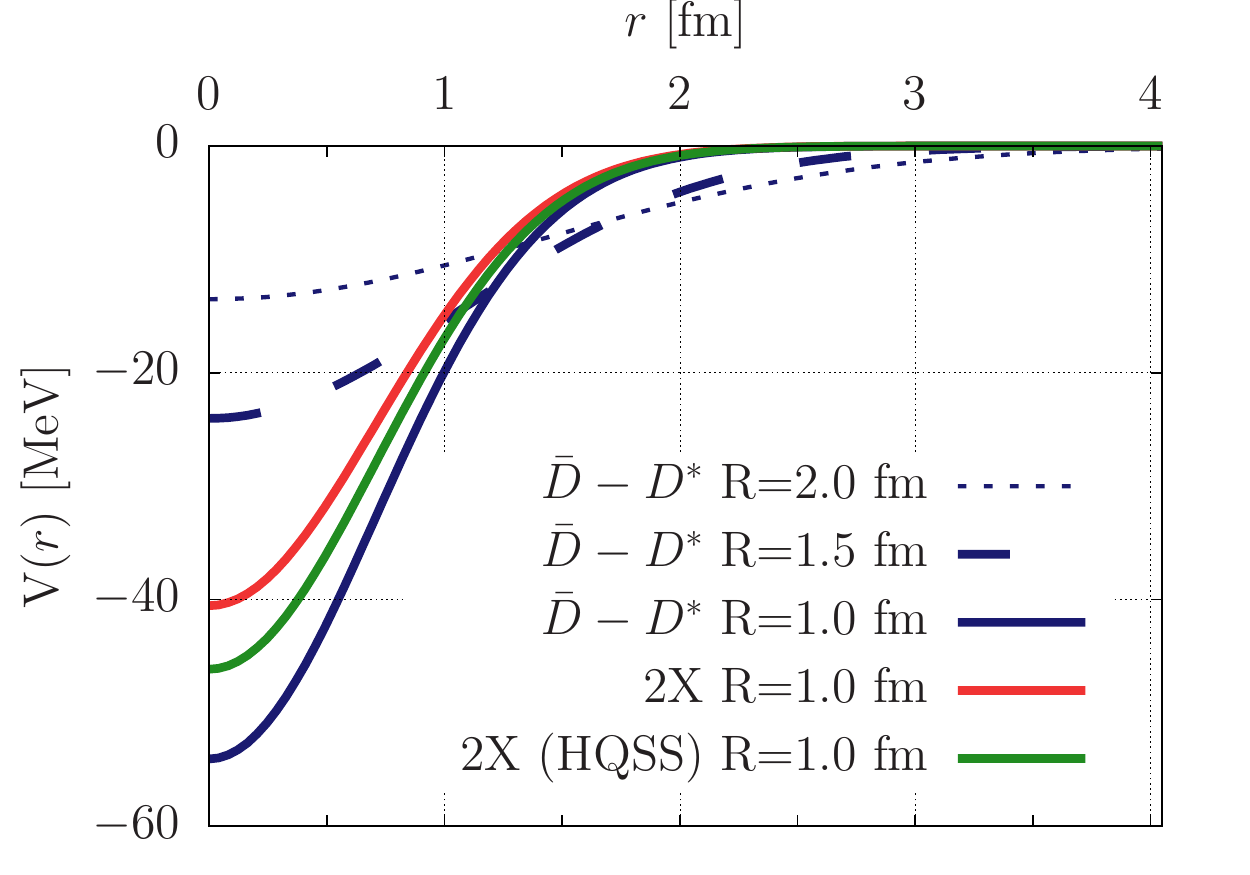} 
\caption{Regularized contact-range $\bar{D}-D^*$ potential
  \eqref{eq:delta-regularization}~ with strengths $C(R_c)$ renormalized
  to a $X$ molecule at threshold for cutoff radii
  $R_c = 1.0, 1.5, 2.0\,{\rm fm}$ (solid, dashed, dotted blue lines),
  \ie, the depth is determined by the coupling strength
  \eqref{eq:delta-regularization}~ and changes with $R_c$
  in order to generate the same diverging two-body scattering length
  independent of $R_c$.
  The effective meson-antimeson interactions for $R_c = 1.0\,{\rm fm}$
    in the four-body 2$X$ system (see \eqref{eq:Vav_2}) are shown
    in red for comparison. The enhancement of the potential
    due to HQSS is shown in green (see \eqref{eq:Vav_2b}).
  In these two cases, the effective interaction is a factor of $0.75$ and
    $0.854$ weaker to that of the $\bar{D}-D^*$ potential
    owing to the existence of non-interacting
    meson-antimeson pairs in the multi-meson wavefunctions
    (see \eqref{eq:X21} to \eqref{eq:Vav_2b}).
}
\label{fig:potential}
\end{figure}
We treat the above-mentioned ``polymers'' as a non-relativistic few-body problem.
The charm meson and antimesons comprising these ``polymers''
have a ground ($D$/$\bar{D}$) and excited ($D^*$/$\bar{D}^*$)
state.
Their isospin $I=1/2$ discriminates between neutral and charged states.
Because of their mass difference, we will only consider
the neutral mesons which dominate the 
$X$ wave function tail.
In the unitary limit,
we have to consider only {\it resonant}\footnote{Two particles interact resonantly if their $S$-wave scattering length diverges and
a two-body bound state with zero energy at threshold exists.} two-body interactions.
Mass and quantum numbers of the $X$ from the molecular perspective 
hint towards such a resonant behaviour
in the $D^{*0} \bar{D}^{0}$ ($1^{++}$) channel
(\ie~ the $X$ channel)~\cite{Braaten:2003he,Voloshin:2003nt},
while HQSS lets us expect the $D^{*0} \bar{D}^{*0}$ ($2^{++}$)
channel to be resonant, too~\cite{Valderrama:2012jv,Nieves:2012tt}.
All other combinations are assumed to be non-resonant and
set to zero. 
Non-resonant interaction pairs would
increase the total binding of the systems slightly. 
Thus they do not alter our conclusions.

To describe the resonant pairs, we employ a contact
two-body potential 
\begin{equation}\label{eq:delta-regularization}
  V(\ve{r}; R_c) = C(R_c)\,\delta^{(3)}(\ve{r}; R_c) \, ,
\end{equation}
regularized with the Gaussian cutoff function
\begin{equation}\label{eq:gauss-regularization}
  \delta^{(3)}(\ve{r}; R_c) = \frac{e^{-(r/R_c)^2}}{\pi^{3/2} R_c^3} \,.\nonumber
\end{equation}
The cutoff-radius $R_c$ and coupling constant $C(R_c)$ are
calibrated (renormalized) to the mass of the $X(3872)$:
\begin{eqnarray}
  M_X =  m(D^0) + m(D^{*0}) - B_X \, .
\end{eqnarray}
Here, $m(D^0)$, $m(D^{*0})$ denote the masses of $D^0$ and $D^{*0}$ mesons, respectively,
and the binding energy $B_X$ is positive for a stable state.
As we are interested in the universal properties of charmed meson clusters,
we will set $B_X = 0$ in accordance
with the current experimental value $B_X = -0.01 \pm 0.18 \,{\rm MeV}$
putting it slightly above threshold~\cite{Zyla:2020zbs}.
The potential underlying this threshold state is attractive, \ie~$C(R_c) < 0$, and
would bind the system if increased by an infinitesimal amount.
The contact-range potential is visualized in fig.~\ref{fig:potential}
for $R_c = 1.0, 1.5, 2.0\,{\rm fm}$, where the relation between an
interaction of shorter range with a larger coupling (\ie~a deeper potential),
as to ensure that the $D^{*0} \bar{D}^0$ bound state is always
at threshold, is apparent.
When solving the Schr\"odinger equation with this potential \eqref{eq:delta-regularization},
we will consider all the mesons to have an identical mass of $m=1933.29$ MeV,
\ie~twice the reduced mass of the $D^{*0} \bar{D}^0$ pair within the $X$
because it is
the most important resonant interaction
in the multi$-X$ systems (see derivation
of \eqref{eq:Vfull}-\eqref{eq:Vav_4} below). 
Corrections are deemed to be subleading and without impact
on the qualitative description of the many-particle states.

Renormalized predictions, in principle, require
that observables are cutoff independent.
We will show below that hexamer and octamer binding energies do not
conform with this demand as they Thomas-collapse if $R_c \to 0$.
In few-body systems, this type of divergence can be renormalized via
an additional three-boson datum~\cite{Bedaque:1998kg,Bedaque:1999ve}~which, 
as of now, is unavailable in the few-$X$ sector.

Despite this obstacle, we can obtain information about the existence of
bound states and estimates of their binding energies.
To this end, we choose a cutoff range near the theory's
breakdown scale which is determined by the longest omitted
short-range component of the interaction.
This missing component is in case of the $X$ the charged channel\footnote{
Pion effects are na\"ively expected to enter perturbatively at subleading orders~\cite{Fleming:2007rp}, suggesting leading-order predictions which are
indistinguishable in pionfull and pionless treatments. Analogous to the conjectured effect of the charged components of the $D$'s,
the inclusion of pions is expected to change the breakdown scale of the theory and
generates a finite-range interaction which
adds to the attraction in larger clusters instead of their disintegration.}, \ie~
the $D^{*+}D^{-}$ component of the $X$ wave
function~\cite{Gamermann:2009fv,Gamermann:2009uq}.
The characteristic momentum scale of the charged channel is
$M_{\rm ch} \simeq 125\,{\rm MeV}$.
It is sensible to expect a cutoff in the vicinity of
$M_{\rm ch} R_c \sim 1$ for which the three-body counterterm vanishes,
and that it remains numerically small within some interval around it.
This smallness suffices to foresee that their inclusion would have no
effect on the character of a state: bound will remain bound,
resonance will remain resonance, {\it etc.}.
Hence, a bound state found within a cutoff
range around $M_{\rm ch} R_c \sim 1$ (specifically, we chose $R_c = 1.0-2.0\,{\rm fm}$)
can reliably be considered a renormalized prediction which will not change character
even with the proper calibration of a collapse-preventing counterterm.

\vspace{2mm}
\textbf{Interaction between meson pairs: }
We exemplify the few-$X$ calculations with a detailed discussion
of the four-body, \ie~two-$X$ problem.
First, we treat the $X$ as a pure $D^{*0} \bar{D}^0$ two-body system.
This approximation disregards the shorter-range $D^{*+} \bar{D}^-$ component
and assumes the $X$ wave function to be
\begin{eqnarray}\label{eq:Xstate}
  \Psi_X = \phi_X (\ve{r})\,\frac{1}{\sqrt{2}}\,\left[ 
  | D^0 \bar{D}^{*0} \rangle + | D^{*0} \bar{D}^0 \rangle \, \right]\;\;,
\end{eqnarray}
with the spatial two-body wave function $\phi_X(\ve{r})$.
The charm meson-antimeson potential in the $X$ channel is defined
for the linear combination $D^{*0} \bar{D}^0 + D^0 \bar{D}^{*0}$
(the positive C-parity combination).
It is practical to use a Fock representation of the potential:
\begin{gather}\label{eq:Vfull}
  V_X(\ve{r}; R_c) =\\
   \frac{V_D(\ve{r}; R_c)}{2} \, \left[
    | D^0 \bar{D}^{*0} \rangle\,\langle D^0 \bar{D}^{*0} | +
    | D^{*0} \bar{D}^{0} \rangle\,\langle D^{*0} \bar{D}^{0} |
    \right]  \nonumber \\
  + \frac{V_E(\ve{r}; R_c)}{2} \, \left[
    | D^0 \bar{D}^{*0} \rangle\,\langle D^{*0} \bar{D}^{0} | +
    | D^{*0} \bar{D}^{0} \rangle\,\langle D^{0} \bar{D}^{*0} |
    \right] \, , \nonumber 
\end{gather}
with a direct ($V_D$) and an exchange term ($V_E$)
which combine to the potential in the $X$ channel,
$V_X = V_D + V_E$.
As no negative C-parity partner of the $X$ has been found yet,
we assume $|V_D + V_E| \gg |V_D - V_E|$.
Moreover, the isospin-breaking decays of the $X$~\cite{Gamermann:2009fv,Gamermann:2009uq}
allow access to $V_E$ and corroborate this inequality~\cite{HidalgoDuque:2012pq}.
Hence, we use the approximation $V_D = V_E = \frac{1}{2} V$ and 
\begin{gather}\label{eq:Vaverage}
  V_X =
  \frac{V}{2} \left[|D^0\bar{D}^{*0}\rangle+|D^{*0}\bar{D}^0\rangle\right]\left[\langle D^0\bar{D}^{*0}|+\langle D^{*0}\bar{D}^0|\right] \, ,
\end{gather}
where the $(\vec{r}, R_c)$ dependence of the potential
has been dropped to improve readability.

The two-$X$ tetramer contains in principle the six possible permutations of
the $| D^0 {D}^{0} \bar{D}^{*0} \bar{D}^{*0} \rangle$ state that result
from exchanging ground- and excited-state mesons
(we assume the spins of all the $D^{*0}$/$\bar{D}^{*0}$ mesons/antimesons
to point in the same direction).
However, these permutations are further constrained by symmetries, as we require
(i) positive C-parity (\ie, invariance {\it wrt.}~the exchange of particles and antiparticles), and (ii)
$D^{0}$ and $D^{*0}$ to obey Bose statistics which we realize with
symmetric internal and spatial wave-function components as they are expected
to provide the majority of the attraction
(\ie, symmetric combinations of $D^{0} D^{*0}$ and $D^{*0} D^0$~\footnote{Antisymmetric combinations -- the nuclear analog are proton-proton or neutron-neutron spin-1 contributions to, \eg~${}^4$He -- demand an odd angular momentum with a perturbatively
small effect in the leading-order framework employed in this work.}).
This reduces the number of relevant states from six to two:
\begin{eqnarray}
  | 1 \rangle &=&
  \frac{| D^0 D^{*0} \bar{D}^{*0} \bar{D}^{0} \rangle + | D^{*0} 
    D^0 \bar{D}^0 \bar{D}^{*0} \rangle
  }{\sqrt{2}}\, , \label{eq:X21}\\
 | 2 \rangle &=&
   \frac{| D^0 D^{*0}   \rangle + | D^{*0} {D}^0 \rangle
   }{\sqrt{2}}\, \frac{| \bar{D}^{*0}  \bar{D}^{0} \rangle
     + | \bar{D}^0 \bar{D}^{*0} \rangle
  }{\sqrt{2}}
  \, . \label{eq:X22}
\end{eqnarray}
In this basis, the potential reads (insert \eqref{eq:X21} and \eqref{eq:X22} in \eqref{eq:Vaverage})
\begin{eqnarray}\label{eq:X12vbar}
  \sum_{ij} V_X(\ve{r}_{ij}; R_c)
  \begin{pmatrix}
    | 1 \rangle \\
    | 2 \rangle 
  \end{pmatrix} =
  \begin{pmatrix}
    2 \bar{V} & \sqrt{2} \bar{V} \\
    \sqrt{2} \bar{V} & \bar{V}
  \end{pmatrix} \,
  \begin{pmatrix}
    | 1 \rangle \\
    | 2 \rangle 
  \end{pmatrix}
  \, ,
\end{eqnarray}
where $\bar{V}$ represents the average of the potential for all resonant pairs.
Considering that $V_X$ involves particle-antiparticle interactions only,
and the same ordering as in $|1\rangle$ and $|2\rangle$ (\ie,
indexing particles before antiparticles):
\begin{equation}\label{eq:Vbar}
\bar{V} = \frac{1}{4}\,\left[V(\ve{r}_{13}) + V(\ve{r}_{14}) + V(\ve{r}_{23}) + V(\ve{r}_{24})\right]\,.
\end{equation}
The diagonalization of \eqref{eq:X12vbar} yields
\begin{eqnarray}
  \sum_{ij} V_X(\ve{r}_{ij}; R_c)\, | X_2 \rangle  = 3 \bar{V} \,| X_2 \rangle
  \, ,
  \label{eq:Vav_2}
\end{eqnarray}
as the most attractive configuration, with \mbox{$|X_2 \rangle = \sqrt{\tfrac{2}{3}} | 1 \rangle + \sqrt{\tfrac{1}{3}} | 2 \rangle$} being
a four-meson eigenstate of $\sum V_X$.
The original coupled-channel problem has thereby been recast
into a single-channel form.

The steps detailed above for the tetramer can be straightforwardly applied
to the hexamer and octamer.
The six-body case comprises, in principle, 20 possible permutations of
the $| D^0 {D}^{0} D^{0} \bar{D}^{*0} \bar{D}^{*0} \bar{D}^{*0} \rangle$ state,
which are reduced to two states by symmetry constraints.
In the eight-body case, there are 70 possible permutations of the
$| D^0 {D}^{0} D^{0} D^{0} \bar{D}^{*0} \bar{D}^{*0} \bar{D}^{*0} \bar{D}^{*0} \rangle$ state, which are reduced to three symmetric ones.
The potential can be diagonalized, as before in the four-body case, leading to a
series of eigenvalues and eigenvectors of which
the most attractive configurations are
\begin{eqnarray}
  \label{eq:Vav_3}
  \sum_{ij} V_X(\ve{r}; R_c)\, | X_3 \rangle  &=& 6 \bar{V} \,| X_3 \rangle \, ,
  \\
  \label{eq:Vav_4}
  \sum_{ij} V_X(\ve{r}; R_c)\, | X_4 \rangle  &=& 10 \bar{V} \,| X_4 \rangle \,.
\end{eqnarray}
Again, $\bar{V}$ represents the average of the potential experienced
by the interacting pairs, while $| X_3 \rangle$, $| X_4 \rangle$ are
the eigenvectors in the internal space of the interaction
that correspond to the most attractive configuration.%

In order to analyze the effect of HQSS on our predictions,
we modify the above-derived interaction.
First, we infer from HQSS a potential
in the $D^{*0} \bar{D}^{*0}$ ($2^{++}$) channel
identical to that in the $X$ channel.
Note the approximate character of this symmetry and
the resulting hypothetical nature of the $2^{++}$
partner of the $X$.
The HQSS extension of the two-body potential
$V_X$ of \eqref{eq:Vaverage}~ is
\begin{eqnarray}\label{eq:VaverageHQSS}
  V_X^{\rm HQSS} &=& V_X + V\,
    | D^{*0} \bar{D}^{*0} \rangle\,\langle D^{*0} \bar{D}^{*0} | \,.
\end{eqnarray}
Coupling between the $1^{++}$ and the $2^{++}$ channels is precluded in the
two-body sector but, nevertheless, these transitions become possible in
the few-$X$ sector, where these states appear as intermediate structures
in the wave function.

We use the four-body case once again to exemplify how the additional interaction term leads to more attraction
than expected earlier. In the basis \eqref{eq:X21}, \eqref{eq:X22}, the potential \eqref{eq:VaverageHQSS}~ now reads
\begin{align}
  \sum_{ij} V_X^{\rm HQSS}(\ve{r}; R_c)
  \begin{pmatrix}
    | 1 \rangle \\
    | 2 \rangle 
  \end{pmatrix} =
  \begin{pmatrix}
    2 \bar{V} & \sqrt{2} \bar{V} \\
    \sqrt{2} \bar{V} & 2 \bar{V}
  \end{pmatrix} \,
  \begin{pmatrix}
    | 1 \rangle \\
    | 2 \rangle 
  \end{pmatrix}
  \, ,
\end{align}
whose diagonalization gives
\begin{eqnarray}
  \sum_{ij} V_X^{\text{HQSS}}(\ve{r}; R_c)\, | X_2' \rangle  =
  (2+\sqrt{2}) \bar{V} \,| X_2' \rangle
  \, , \label{eq:Vav_2b}
\end{eqnarray}
with the more attractive eigenvalue $(2+\sqrt{2})~\bar{V}$ and eigenvector $|X_2' \rangle = (| 1 \rangle + | 2 \rangle) / \sqrt{2}$.
For the six- and eight-body systems, the same assumptions and symmetries result in
\begin{eqnarray}
  \sum_{ij} V_X^{\rm HQSS}(\ve{r}; R_c)\, | X_3' \rangle  &=&
  \frac{1}{2}(11+\sqrt{13}) \bar{V} \,| X_3' \rangle \, , \label{eq:Vav_3b} \\
  \sum_{ij} V_X^{\rm HQSS}(\ve{r}; R_c)\, | X_4' \rangle  &=&
    \left(8+\sqrt{22}\right)\, \bar{V} \,| X_4' \rangle \label{eq:Vav_4b} \, , 
\end{eqnarray}
with eigenvectors $|X_3' \rangle$ and $|X_4' \rangle$ corresponding
to configurations in which the potential is most attractive.

In both cases, with and without HQSS, the spectrum of a system
composed of $N/2$ $X$ particles is given by the Schr\"odinger equation 
\begin{align}\label{eq:NB}
\left(- \sum^N_{i<j} \frac{\hbar^2}{2\,m}\left(\ve{\nabla}_{ij}\right)^2
+\eta  
\sum_{{1 \leq i \leq N/2 \atop  (N/2 +1) \leq j\leq {N}}}
  V(\ve{r}_{ij}) \right) \phi_N(\ve{r})\nonumber\\
= E~\phi_N(\ve{r}) \, ,
\end{align}
with $\phi_N(\ve{r})=\left\langle\ve{r}_1,\ve{r}_2,\cdots,\ve{r}_N| {(N/2)X}\right\rangle$, $\eta$ the eigenvalues of \eqref{eq:X12vbar}~ ($\eta= \{$3/4, 6/9, 10/16$\}$ respectively for $2X$, $3X$, and $4X$),
$\ve{r}_{ij}=\ve{r}_i-\ve{r}_j$ with the index $i$($j$) representing a charm
meson (antimeson), where we have indexed the particles first and
then the antiparticles, and with $m$ being twice the reduced
mass of the $D^0 D^{*0}$ system, \ie~$m\approx1933~$MeV.
Finally, $E$ refers to the energy of the $N$-body system, where we are interested in bound states ($B=-E > 0$).

In practice, we solve the Schr\"odinger equation
with the Stochastic-Variational Method (SVM~\cite{Suzuki:1631377,Suzuki:1998bn}).
In our implementation, this method expands the wave function in correlated Gaussian functions ($(N-1)\times(N-1)$ relative Jacobi coordinates),
with a non-zero interaction between the relevant pairs (meson-antimeson).
We abstain from an explicit symmetrization of the spatial wave function, \ie, we do not project onto $L=0$ and assume
that the central and parity-preserving character of the potential will produce the energetically favourable symmetric ground states
in the course of the variational optimization.

\vspace{2mm}
\textbf{Results and conclusions: }
\begin{figure}
\centering
\includegraphics[width=\linewidth]{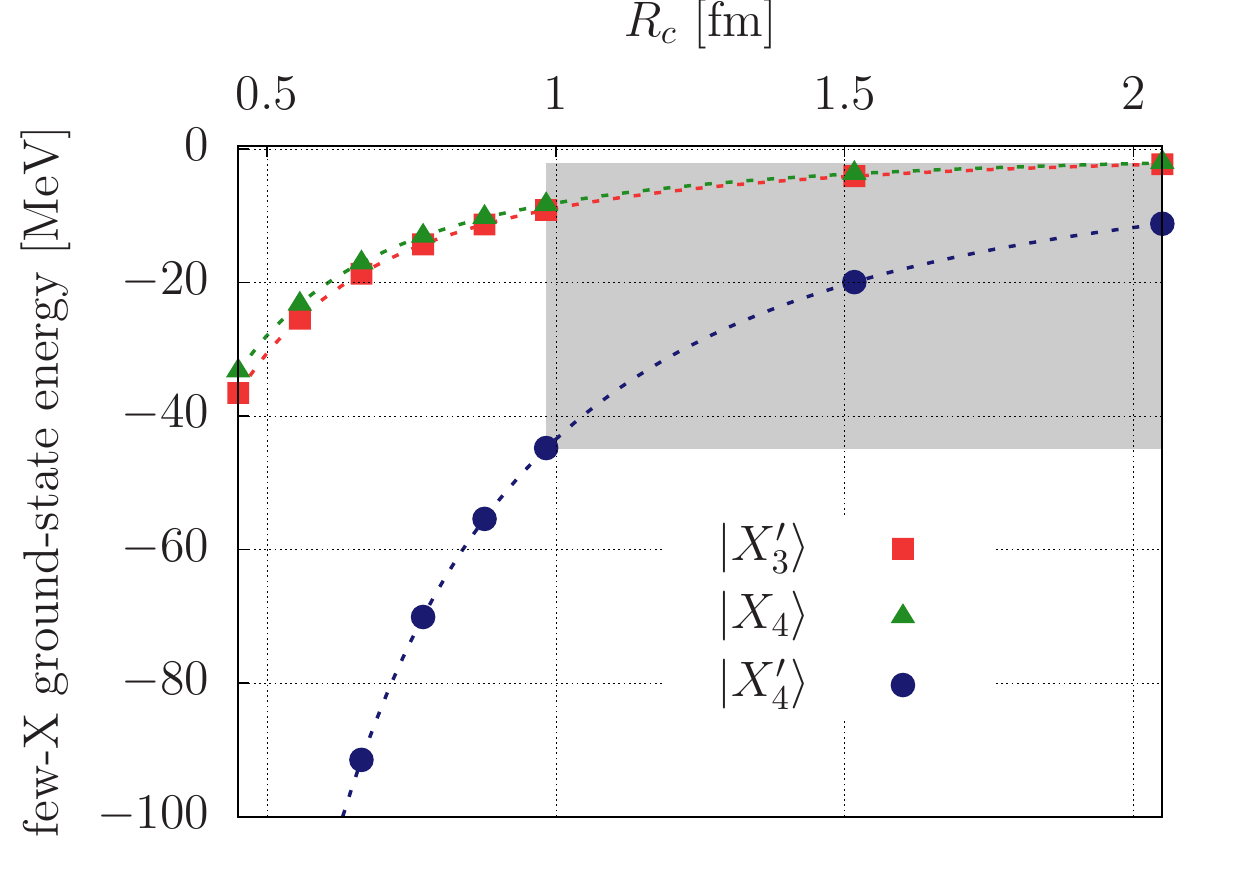} 
\caption{
  Cutoff-radius dependence of the ground-state binding energies of
  few-$X$ systems. With a resonant meson-antimeson interaction
  in the $X$ channel and the $J^{PC}=2^{++}$ partner channel,
  3$X'$ --~a 6-meson state~-- (red square, dotted) and 4$X'$ --~an 8-meson state~-- (blue square, dotted) clusters are bound.
  Solely with a resonant $X$-channel interaction, only the 4$X$ (blue circle, dashed)
  is bound.
  The binding energies are proportional to $1/R_c^2$ (dashed/dotted lines),
  and indicate a Thomas collapse of the systems.
  The ensuing counterterm(s) are expected to vanish within the gray-shaded area,
  while the total $R_c$ range spans from the typical hadron size up to a scale
  set by the expected charged components of the $X$.
  }
\label{fig:binding}
\end{figure}
Assuming the charm meson-antimeson interaction in the $X$-channel
to dominate, \ie, with the average interactions
\eqref{eq:Vav_2}, \eqref{eq:Vav_3}, and \eqref{eq:Vav_4},
we find solutions to \eqref{eq:NB}~ of the four-body ($2X$) and six-body ($3X$)
systems to be unbound.
Adding another $X$, we predict the eight-body ($4X$) system bound
with $B_{4X} > 2.08\,{\rm MeV}$.
Including the attraction induced by HQSS, \ie, the interactions in \eqref{eq:Vav_2b}, \eqref{eq:Vav_3b}, and \eqref{eq:Vav_4b},  the eight-body binding energy
increases to \mbox{$B_{4X}^{HQSS} > 11.21\,{\rm MeV}$.}
Furthermore, the six-body system becomes bound with $B_{3X} > 2.29\,{\rm MeV}$.
These results represent sensible lower bounds for the binding energies of
the respective systems obtained
at a regularization scale of about $2\,{\rm fm}$,
a value deemed soft enough for an attractive three-body
counterterm.
Furthermore, any attraction from the non-resonant mesonic interactions
(set to zero in our calculations) is expected to increase binding energies.

As alluded to in the introduction, this appearance of few-body clusters bound by a few MeV 
as a result of a resonant two-body system with close-to zero binding energy is not unprecedented.
In comparison with those
universal $A = 4,5,6$ bosonic clusters found~\cite{vonStecher:2011zz,Bazak:2016wxm}
attached to $A=3$ Efimov states, the constraint of a resonant interaction in the $X$-channel, only,
amounts to an $N$-body problem with each particle interacting only with $N/2$ particles.
Furthermore, the strength of this interaction is reduced and is no longer
resonant. To be explicit, instead of 6(28) resonant interaction pairs in a system of
4(8) bosons, the limitation to resonant interaction only in the $1^{++}$ channel 
yields a 4(8) equal-mass boson problem with 4(16) non-resonant interaction pairs.
Multi-$X$ boundstates are thus not expected to expose molecular-$X$ behaviour.
Therefore, they should be approached,
as done in this study, as multi-$D$/$\bar{D}$ systems.
We also notice that three $X$ bosons do not show a Efimov spectrum.

In figure \ref{fig:binding}, we show the
regulator dependence of
the binding energies as a signature of the Thomas collapse.
Originally this collapse is expected for one-channel systems of
identical particles in the zero-range limit~\cite{Thomas:1935zz}.
Here, we demonstrate the occurrence of the collapse for a more complex system
with more than one channel and where a certain number of
interaction pairs have been removed.
A range of cutoffs over which the effect of the unenforced renormalization
condition (\eg, the canonical three-body counterterm) is expected to vanish
is marked in the figure (gray area).

Another effect of the reduction of resonantly interacting pairs found here
is the cutoff-independent ratio between $4X$ and $3X$ energies, $B_4/B_3\sim4.9$.
Compared with the ratio found in~\cite{Deltuva:2010xd}
and~\cite{Carlson:2017txq}, $B_4/B_3\sim4.6$,
we conclude that reducing the number of interacting pairs
widens the gap between the energy of $N$- and $(N+1)$-boson systems. 
However, a single counterterm should still suffice to renormalize both systems.
Interpreted more generally, this study hints towards new universal systems in which part of the resonant interactions
are amplified by the presence of more two-body channels or totally removed by symmetry effects.
Consequences of deviations from universality and the effect of multiple open two-body channels on universal ratios
are problems beyond the scope of this work.

\begin{figure}
\centering
\includegraphics[width=\linewidth]{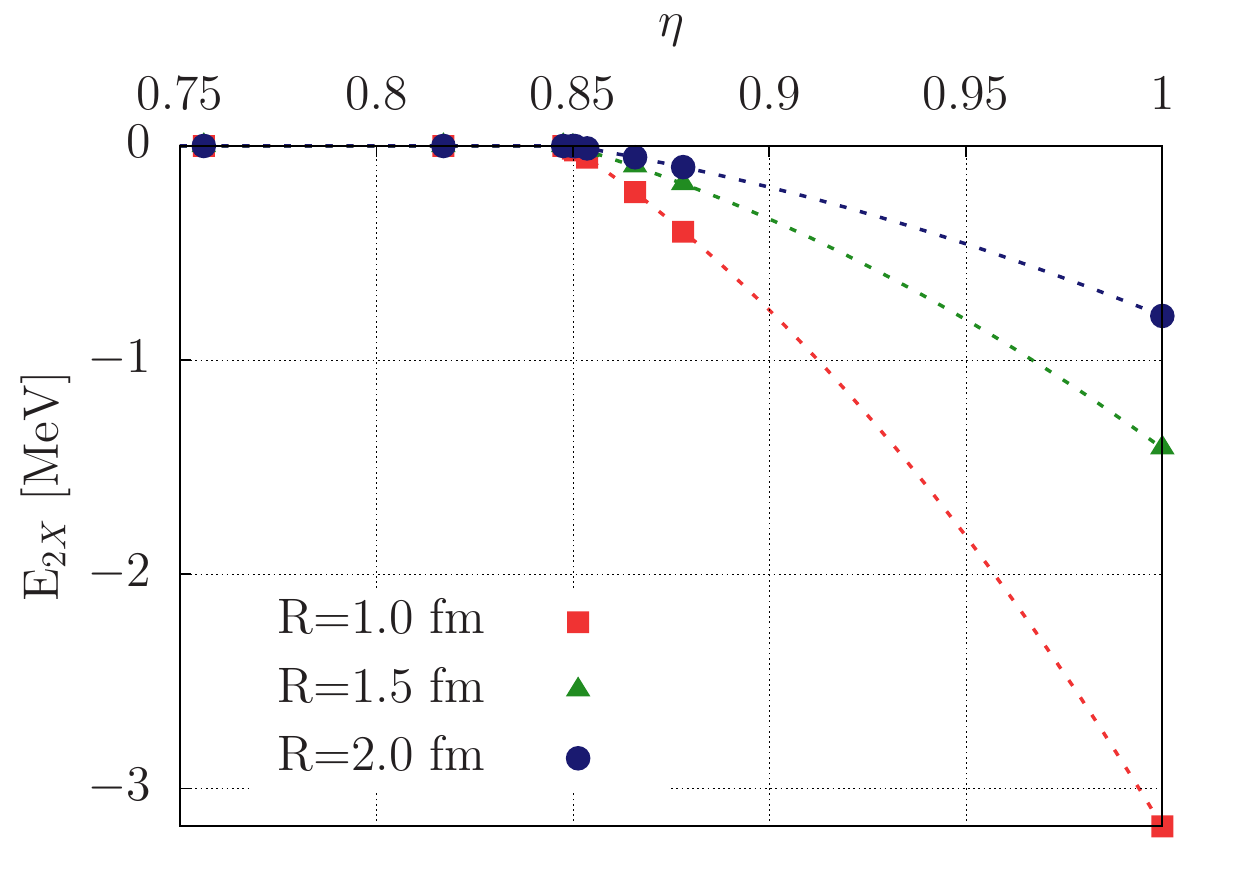} 
\caption{
    Ground-state energy of the four-body 2$X$ system as a function of
    the dimensionless coupling strength $\eta$ (see \eqref{eq:NB}).
    A bound state forms for $\eta \geq \eta^* = 0.876(1)$.
    If only the $X$-channel interaction (\ie~the $D^{*0} \bar{D}^0$ ($1^{++}$) 
    system) is considered to be resonant then $\eta = 0.75$,
    which is insufficient for the formation of a 2$X$ tetramer.
    If the $D^{*0} \bar{D}^{*0}$ ($2^{++}$) channel is also considered
    (see \eqref{eq:Vav_2c}) then $\eta = 0.854$, pretty close to $\eta^*$
    Finally, for an attractive $D^{0} \bar{D}^0$ ($0^{++}$) interaction
    with at least $17.4(8)\%$ of the $X$ channel strength,
    $\eta$ surpasses the critical value and the $2X$ binds.
    The dependence of the bound-state energy on $\eta$ fits a parabola
    for all considered cutoffs (dotted lines).
  }
\label{fig:eta}
\end{figure}

Finally, we revisit the $X_2$ and assess the conditions for its binding.
The heavy-quark content of the $X_2$ is $cc\bar{c}\bar{c}$, a
double charm-anticharm content which was first observed
when detecting $\Xi_{cc}^{++}$~\cite{Aaij:2017ueg}
and more recently the fully charm tetraquark~\cite{Aaij:2020fnh}.
These measurements indicate a possible $X_2$ discovery in the near future.
In this regard, it is interesting to notice that the $D^0\bar{D}^0$ interaction
have been theorized to be attractive and strong enough as to even support
a bound state~\cite{Zhang:2006ix,Gamermann:2006nm,Nieves:2012tt}.
As this interaction is not connected to the $V_X$ potential via HQSS,
its strength is unknown. Its structure, however, can be included it our
framework by refining \eqref{eq:VaverageHQSS} as follows:
\begin{eqnarray}
  \tilde{V}^{\rm HQSS}_X = V^{\rm HQSS}_X +
  \lambda \, V | D^0 \bar{D}^0 \rangle \langle D^0 \bar{D}^0 |
  \, .
\end{eqnarray}
Here, $\lambda$ parametrizes the relative strength of the interaction
in the $D^0 \bar{D}^0$ ($0^{++}$) channel with respect to the $X$.
The diagonalization of this modified potential (see \eqref{eq:Vav_2}) is
\begin{eqnarray}
  \sum_{ij} \tilde{V}_X^{\text{HQSS}}(\ve{r}; R_c)\, | X_2'' \rangle  &=& 
  (2 + \frac{\lambda}{2} +
  \sqrt{2 + \frac{\lambda^2}{4}}) \bar{V} \,| X_2'' \rangle
  \nonumber \\
  &=& 4 \, \eta(\lambda)\, \bar{V}\,| X_2'' \rangle\;\;.
  \label{eq:Vav_2c}
\end{eqnarray}
Compared with \eqref{eq:NB}, the strength of the potential $\eta$ is now
a function of the additional attraction parameter $\lambda$.
We find numerically that if $\eta \geq \eta^*=0.876(1)$ the four-body (2$X$) system
binds (see Fig.~\ref{fig:eta}). This critical strength is approximately
constant for all considered cutoffs $R_c = 1.0, 1.5, 2.0\,{\rm fm}$.
An analysis and explanation of this phenomenon remains beyond this article.
The condition $\eta \geq \eta^*$ is equivalent to $\lambda \geq \lambda^* = 0.174(8)$, \ie~,
the 2$X$ will bind if an additional $(>20\%)$ $D^0 \bar{D}^0$ attraction is provided.
As bound states of the
$D^0 \bar{D}^0$ system are a conceivable scenario~\cite{Zhang:2006ix,Gamermann:2006nm,Nieves:2012tt,Prelovsek:2020eiw,Dong:2021juy}, 
the enhanced attraction and the ensuing bound 2$X$ tetramer are plausible.

In summary, we have shown how the substructure of a unitary dimer
-- the $X$ -- affects the spectrum of its cluster states.
This spectrum differs from the one predicted %
for point-like bosons in the unitary limit~\cite{Carlson:2017txq}
in an intriguing aspect.
Namely, under certain assumptions about the meson-antimeson interaction,
the $X$ cluster states realize a novel generalization of
Borromean/Brunnian systems.
Regardless of the enormous practical difficulties which hamper
an experimental (or numerical, in the lattice) verification of
our conjectures (double charm-anticharm production has only been
recently achieved~\cite{Aaij:2017ueg,Aaij:2020fnh}),
we deem the exposition of the mechanism which ``delays''
the formation of bound structures -- the onset of binding with 4$X$ and 3$X$,
but not necessarily with 2$X$ under the assumptions we made --
as a noteworthy result of the above.
Yet in this later case the 2$X$ tetramer will bind provided we include
  a weakly attractive $D^0 \bar{D}^0$ interaction in our calculations
  of about $20\%$ the strength of the X-channel potential.
  In view of previous conjectures about a possible $D^0 \bar{D}^0$
  molecule~\cite{Zhang:2006ix,Gamermann:2006nm,Nieves:2012tt},
  this condition might very well be met in the real world.
  If this is to be the case, the 2$X$ tetramer will be experimentally
  accessible in the near future.

\vspace{2mm}
\textbf{Acknowledgments: }
We thank Feng-Kun Guo (Institute of Theoretical Physics in Beijing, CAS) and Michael C. Birse (The University of Manchester) for valuable comments and discussions.
This work is partly supported by the National Natural Science Foundation
of China under Grants No.  11735003, No. 11975041 and the Fundamental
Research Funds for the Central Universities.
L. Contessi acknowledges the support and the framework of the "Espace de Structure et de réactions Nucléaires Théorique" (ESNT, http://esnt.cea.fr ) at CEA.
M.P.V. is supported by the Thousand Talents Plan for Young Professionals and would like to thank the IJCLab of Orsay for its long-term hospitality.

\bibliography{tripleX2.0arxiv.bib}

\end{document}